\documentclass[fleqn]{annalen}
\newcommand{\bm}[1]{{\mbox{\boldmath$#1$}}}
\begin{document}
\newcommand{\volume}{11}             %sets current volume,
\newcommand{\xyear}{2000}            %sets year in header
\newcommand{\issue}{5}               %sets current issue,
\newcommand{\recdate}{15 November 1999}  %sets received date,
\newcommand{\revdate}{dd.mm.yyyy}    %sets revised date,     
\newcommand{\revnum}{0}              %number of revisions,
\newcommand{\accdate}{dd.mm.yyyy}    %sets accepted date,
\newcommand{\coeditor}{ue}           %sets (co)editor,
\newcommand{\firstpage}{000}         %first page number,  
\newcommand{\lastpage}{000}          %last page number,
\setcounter{page}{\firstpage}        %sets page counter to first page number 
\newcommand{\keywords}{General Relativity, the Doppler effect, Astrometry}
\newcommand{\PACS}{04.20.Cv, 04.25.-g, 04.80.-y}
\newcommand{\shorttitle}
{S.M. Kopeikin, Lorentz Covariant Theory of Precise Doppler Measurements}
\title{Lorentz Covariant Theory of Precise Doppler Measurements}
\author{Sergei M. Kopeikin$^{1,2}$}
\newcommand{\address}
{$^{1}$ Dept. of Physics \& Astronomy, UMC, Phys. Build. 223, Columbia, 
MO 65211, USA\\
$^{2}$ On leave: Astro Space Center, FIAN, Moscow, Russia}
\newcommand{\email}{\tt KopeikinS@missouri.edu} 
\maketitle
\begin{abstract}
The Lorentz covariant theory of precise Doppler measurements (PDM) based on
the retarded \textit{Li\'{e}nard-Wiechert} solution of the Einstein
equations is described. An exact solution of equations of light propagation in the
field of arbitrary moving bodies, which drastically
extends the range of applicability of the new theory of PDM, is obtained.
An explicit formula for the gravitational shift of
frequency is given in analytic form. The limiting cases of the
Doppler observations in gravitational lensing and of the spacecraft's
Doppler tracking are described in more
detail. We also present the post-Newtonian theory of the PDM 
developed for searching relativistic effects in close optical binaries and
massive planetary systems.
\end{abstract}

\section{Background}

Detection of low-frequency gravitational waves and verification of General
Relativity in the Solar system Doppler tracking experiments, as well as
searching for anysotropies in the cosmic background radiation (CMB) 
appeal to the development of
the relativistic theory of precise Doppler measurements (PDM). Previous
attempts to create such a theory in framework of General Relativity (GR) 
were based on making use of the
post-Newtonian approximations (PNA) which operate with the instantaneous
values of metric tensor taken on hypersurfaces of constant time. Hence, 
an adequate treatment of propagation of
light rays in the gravitational field of an isolated astronomical system can
be achieved using the PNA only for light-time intervals which do not exceed the
characteristic Keplerian time of the system. If light propagates longer, the
residual terms of the PNA theory of PDM become unmanageable.
Moreover, the question about the instant of time at which positions and
velocities of gravitating bodies deflecting light rays should be fixed to
minimize calculational errors cannot be solved in the framework of the
PNA unambiguously.

A natural way to overcome incompleteness in the PNA approach to
the theory of PDM is to employ the post-Minkowskian approximation (PMA) scheme,
which is Lorentz covariant and does not involve the slow-motion expansions with
respect to powers of $v/c$, where $v$ is the characteristic velocity of the
light-ray-deflecting body, and $c$ is the speed of light. We have succeeded
in developing of such a Lorentz covariant theory of PDM which is based on the
retarded \textit{Li\'{e}nard-Wiechert} solution of the Einstein equations,
linearized with respect to Newton's gravitational constant $G$ \cite{KS99}. 
This makes it possible to obtain a unique exact solution of the
equations of light propagation in the field of arbitrary moving
spherically-symmetric bodies, which drastically extends the range of
applicability of the new theory of PDM. 

\section{Doppler effect in Special and General Relativity}

Special relativistic treatment of the frequency shift is 
well-known. It is based on two facts: 
1) the proper time runs differently for identical clocks moving with different 
velocities; -2) electromagnetic waves propagate along straight lines in 
the flat space-time (see the paper \cite{Ocun} for more details). 

General relativistic formulation of the frequency shift in curved space-time
is more involved. Two definitions of the Doppler shift are used \cite{Synge} -
in terms of energy ($A$) and in terms of frequency ($B$)
\begin{equation}
\label{1} 
(A)\quad\quad\frac{\nu }{\nu _{0}}=\frac{u^{\alpha }\mathcal{K}_{\alpha }}{u_{0}^{\alpha }%
\mathcal{K}_{0\alpha }}\;,\quad\quad\quad\quad\quad\quad\quad (B)\quad\quad
\frac{\nu }{\nu _{0}}=\frac{dT_{0}}{dT}\;,
\end{equation}
where $\nu_0$ and $\nu$ are emitted and observed electromagnetic 
frequencies of light; $T_0$, $u_0^{\alpha }$
and $T$, $u^{\alpha }$ are the proper time and 4-velocity of the source of light 
and observer; 
${\cal K}_{0\alpha}$ and ${\cal K}_{\alpha}$ are null 4-vectors of the light
particle at the points of emission and observation respectively. Despite the 
obvious difference in the two definitions they are identical as
$u^{\alpha}=dx^{\alpha}/dT$, ${\cal K}_{\alpha}=
\partial\varphi/\partial x^{\alpha}$, and the phase of electromagnetic wave
$\varphi$ is constant along the light rays. In GR there does not exist 
a parallel transport of 4-vectors in the sense of the flat space-time.
Instead of that one has to integrate equations of light propagation to 
connect physical
quantities at the points of emission and observation of light.  
 
The integration of null geodesic equations in the case of space-times 
possesing symmetries was known for a long time and extensivly
used in astronomical practice \cite{Brumberg}. 
However, the most interesting astronomical phenomena in the propagation 
of light rays in curved space-time are caused by small, time-dependent 
perturbations of the background geometry. Usually, the 1st PNA 
in the relativistic N-body problem with fixed or uniformly moving 
bodies was applied to consider effects of such perturbations \cite{KK}. 
Unfortunately, this 
approximation works properly if, and only if, time of propagation of light is much
shorter than the characteristic Keplerian time of the N-body problem. 
An adequate treatment of the effects in
propagation of light must account for the retardation effects in the
propagation of gravitational field to the point of the field's interaction 
with the electromagnetic wave.  

\section{Predictive Relativistic Mechanics of Photons}

Physically consistent description of gravitational field in the problem of 
propagation of light rays is achieved in the framework of the
post-Minkowskian approximations (PMA) of the Einstein gravitational field equations.
For our purpose one PMA is completely enough. This approximation is linear 
with respect to the universal gravity constant $G$ and gives metric 
tensor as an
algebraic sum of the Minkowski metric $\eta_{\alpha\beta}$ and its
perturbation $h_{\alpha\beta}$ written in terms of 
the {\it Li\'{e}nard-Wiechert} tensor potential. The perturbation 
$h_{\alpha\beta}[t,{\bf x},{\bf x}_a(s),{\bf v}_a(s)]$ is a function of a field 
point $(t,{\bf x})$, masses $m_a$ as well as coordinates ${\bf x}_a(s)$ 
and velocities ${\bf v}_a(s)$
of gravitating bodies taken at the retarded time 
$s=t-\frac{1}{c}|{\bf x}-{\bf x}_a(s)|$.  
Substitution of $h_{\alpha\beta}$ into the equations of the light ray 
geodesics
bring them to the form of to the "retarded-functional differential 
system" (hereafter $G=c=1$)
\begin{equation}
\label{eqs}
\ddot{\mathbf{x}}(t)=\sum_{a=1}^{N}m_a\mathbf{F}_a\left[ \mathbf{x}(t),\dot{%
\mathbf{x}}(t),\mathbf{x}_{a}(s),\mathbf{v}_{a}(s)\right] \;,
\end{equation}
because of the dependence of the gravitational light-deflecting force 
${\bf F}_a$  
on the retarded time argument $s$. Studying of such equations belongs 
to the framework of "predictive relativistic mechanics" \cite{Bel}. 

\section{Solution of the light-ray geodesics}

The system (\ref{eqs}) is transformed to a simpler form by making use of 
a specific differential identity \cite{KS99} applied to the force 
${\bf F}_a$ in order to change of the order of operations of taking partial 
derivatives and substitution for the unperturbed light-
ray trajectory. Solution of the system (\ref{eqs}) is, then, achieved by 
performing integrals from ${\bf F}_a$ along the 
light-ray path by making use of transformation from 
the coordinate time $t$ to the retarded time $s$ which 
eliminates the time $t$ from
all of the integrands of any integral. Hence, the 
integrals can be directly performed as soon as the motion of the 
light-ray-deflecting bodies is known. As a result,
the time of light travelling from the point of emission $\bf{x}_0$ to the point of 
observation ${\bf x}$ is obtained in the following form \cite{KS99} 
\begin{equation}
\label{time}
t-t_{0}=|\mathbf{x}-\mathbf{x}_{0}|+
\;2\sum_{a=1}^{N}m_{a}\;{\displaystyle{%
\int_{s_{0}}^{s}}}\frac{[1-\mathbf{k}\cdot \mathbf{v}_{a}(\zeta )]^{2}}{%
\sqrt{1-v_{a}^{2}(\zeta )}}\frac{d\zeta }{t^{*}+\mathbf{k}\cdot \mathbf{x}%
_{a}(\zeta )-\zeta }\;,
\end{equation}
where $t^*$ is a pre-defined constant.
General formula (\ref{time}) is used for calculation of the 
Doppler shift in what follows.

\section{Doppler Effect in Gravitational Lenses}

In the case of a gravitational lens having a total mass ${\cal M}$, 
moving with velocity $V^i$ with respect 
to the chosen coordinate system, and 
possesing a time-dependent quadrupole moment ${\cal I}_{ij}$ 
the gravitational shift of frequency reads \cite{KS99} 
\begin{equation}
\label{grl}
\left( \frac{\delta \nu }{\nu _{0}}\right) _{gr}^{obs}=4
\left( -\mathcal{M}V^{i}\;\hat{\partial%
}_{i}+\frac{1}{2}\dot{\mathcal{I}}^{ij}\hat{\partial}_{ij}\right) \ln |{{%
\mbox{\boldmath$\xi$}}}|
+\frac{r_{0}}{R}\left( \mathbf{v}\cdot {{%
\mbox{\boldmath$\alpha$}}}\right) +\frac{r}{R}\left( \mathbf{v}_{0}\cdot {{%
\mbox{\boldmath$\alpha$}}}\right) \;,
\end{equation}
where 
$r$, $r_0$ are distances from the lens to observer and source of light,
$R\simeq r+r_0$, $\bm{\alpha}$
is the angle of the total deflection of light rays, ${\bf v}$ and ${\bf v}_0$
are velocities of observer and the source of light respectively, 
${\bm{\xi}}=\xi^i$ is the impact parameter of the light ray, 
$\hat{\partial}_i=\partial/\partial\xi^i$.

\section{Spacecraft Doppler Tracking}

Lorentz covariant approach allows to derive general relativistic corrections to
the well-known special relativistic expression for the electromagnetic 
frequency shift in the spacecraft's Doppler tracking. The corrections 
are summarized in the formula \cite{KS99} 
\begin{equation}
\frac{\nu }{\nu _{0}}=\frac{1-\mathbf{k}\cdot \mathbf{v}}{1-\mathbf{k}\cdot 
\mathbf{v}_{0}}\;\left[ \frac{1-v_{0}^{2}}{1-v^{2}}\right] ^{1/2}\;\left[ 
\frac{\mathrm{a}(t_{0})}{\mathrm{a}(t )}\right] ^{1/2}\;\frac{\mathrm{%
b}(t )}{\mathrm{b}(t_{0})}\;,
\end{equation}
which consists of three factors. The first factor is the special
relativistic Doppler shift while 
the second and the third factors are general relativistic corrections. 
They are
rather complicated functions of positions and velocities of observer, 
spacecraft, and light-ray-deflecting masses. Exact expressions for these 
functions can be found in \cite{KS99}. 

\section{Post-Newtonian Theory for Doppler Measurements of Binary Stars}

The determination of velocities of stars from precise Doppler measurements
is described in \cite{KO99}.
We apply successive Lorentz transformations and the relativistic
equation of light propagation to establish the exact treatment of the Doppler
effect in binary systems both in Special and General Relativity theories. As
a result, the Doppler shift is a sum of (1) linear in $c^{-1}$ terms, which
include the ordinary Doppler effect and its variation due to the secular
radial acceleration of the binary with respect to observer; (2) terms
proportional to $c^{-2}$, which include the contributions from the quadratic
Doppler effect due to the relative motion of binary star with respect to
the Solar system, motion of the particle emitting light, 
orbital motion of the star around the binary's barycenter, 
diurnal rotational motion of observer,
and orbital motion of the Earth; and (3) terms
proportional to $c^{-2}$, which include the contributions from redshifts due
to gravitational fields of the star, star's companion, Galaxy, Solar system,
and the Earth. We briefly discuss in the paper \cite{KO99} 
feasibility of practical implementation of these theoretical
results, which crucially depends on further progress in the technique of
precision Doppler measurements. 

\vspace*{0.25cm} \baselineskip=10pt{\small \noindent The author 
thanks G. Neugabauer for constant support and G. Sch\"afer for useful discussions.


\begin{thebibliography}{KS99}
\bibitem[1]{KS99}  S.M. Kopeikin and G. Sch\"{a}fer, Phys. Rev., {\bf D60},
(1999), 124002
\bibitem[2]{Ocun}  L.B. Ocun, K.G. Selivanov, and V.L. Telegdi, 
%e-print physics/9907017, 
submitted to Amer. J. of Phys., (1999)     
\bibitem[3]{Synge} J.L. Synge, {\it Relativity: The General Theory},
North-Holland: Amsterdam, 1971
\bibitem[4]{Brumberg} V.A. Brumberg, {it Essential Relativistic Celestial
Mechanics}, A. Hilger: Bristol, 1991
\bibitem[5]{KK} S.A. Klioner and S.M. Kopeikin, Astron. J. (USA), 
{bf 104}, (1992), 897
\bibitem[6]{Bel} L. Bel, Ann. Inst. H. Poincar{\'e}, {\bf 12}, (1970), 307
\bibitem[7]{KO99} S.M. Kopeikin and L.M. Ozernoy, Astrophys. J., {\bf 523},
(1999), 771  
\end{thebibliography}
\end{document}